# Valence band electronic structure evolution of graphene oxide upon thermal annealing for optoelectronics

**Hisato Yamaguchi\*,[1], Shuichi Ogawa[2], Daiki Watanabe[2], Hideaki Hozumi[2], Yongqian Gao[3], Goki Eda[4,5,6], Cecilia Mattevi[7], Takeshi Fujita[8], Akitaka Yoshigoe[9], Shinji Ishizuka[10], Lyudmyla Adamska[11,12], Takatoshi Yamada[13], Andrew M. Dattelbaum[1], Gautam Gupta[1], Stephen K. Doorn[3], Kirill A. Velizhanin[11], Yuden Teraoka[9], Mingwei Chen[8], Han Htoon[3], Manish Chhowalla[14], Aditya D. Mohite\*\*,[1], and Yuji Takakuwa\*\*\*,[2]**

[1] MPA-11 Materials Synthesis and Integrated Devices (MSID), Materials Physics and Applications (MPA) Division, Mail Stop: K763, Los Alamos National Laboratory (LANL), P.O. Box 1663, Los Alamos, New Mexico 87545, U.S.A.
[2] Institute of Multidisciplinary Research for Advanced Materials (IMRAM), Tohoku University, 2-1-1 Katahira, Aoba-ku, Sendai 980-8577, Japan
[3] Center for Integrated Nanotechnologies (CINT), MPA Division, Mail Stop: K771, LANL, P.O. Box 1663, Los Alamos, New Mexico 87545, U.S.A.
[4] Department of Physics, National University of Singapore, 2 Science Drive 3, Singapore 117542
[5] Department of Chemistry, National University of Singapore, 3 Science Drive 3, Singapore 117543
[6] Graphene Research Centre, National University of Singapore, 6 Science Drive 2, Singapore 117546
[7] Department of Materials, Imperial College London, Exhibition road, London SW7 2AZ, U.K.
[8] WPI Advanced Institute for Materials Research (AIMR), Tohoku University, 2-1-1 Katahira, Aoba-ku, Sendai, Miyagi 980-8577, Japan
[9] Quantum Beam Science Directorate, Japan Atomic Energy Agency, Kouto, Sayo-cho, Sayo-gun, Hyogo 679-5198, Japan
[10] Department of Materials Science and Engineering, Akita National College of Technology, 1-1 Bunkyo-machi, Iigima, Akita 011-8511, Japan
[11] T-1 Physics and Chemistry of Materials, Theoretical Division, Mail Stop: B221, LANL, P.O. Box 1663, Los Alamos, New Mexico 87545, U.S.A.
[12] Center for Nonlinear Studies (CNLS), Theoretical Division, Mail Stop: B258, LANL, P.O. Box 1663, Los Alamos, New Mexico 87545, U.S.A.
[13] Nanotube Research Center, National Institute of Advanced Industrial Science and Technology (AIST), 1-1-1 Umezono, Tsukuba, Ibaraki 305-8568, Japan
[14] Department of Materials Science and Engineering, Rutgers University, 607 Taylor Road Piscataway, New Jersey 08854, U.S.A.

Corresponding Authors: \*e-mail hyamaguchi@lanl.gov
\*\*e-mail amohite@lanl.gov
\*\*\*e-mail takakuwa@tagen.tohoku.ac.jp

We report valence band electronic structure evolution of graphene oxide (GO) upon its thermal reduction. Degree of oxygen functionalization was controlled by annealing temperatures, and an electronic structure evolution was monitored using real-time ultraviolet photoelectron spectroscopy. We observed a drastic increase in density of states around the Fermi level upon thermal annealing at ~600 °C. The result indicates that while there is an apparent band gap for GO prior to a thermal reduction, the gap closes after an annealing around that temperature. This trend of band gap closure was correlated with electrical, chemical, and structural properties to determine a set of GO material properties that is optimal for optoelectronics. The results revealed that annealing at a temperature of ~500 °C leads to the desired properties, demonstrated by a uniform and an order of magnitude enhanced photocurrent map of an individual GO sheet compared to as-synthesized counterpart.

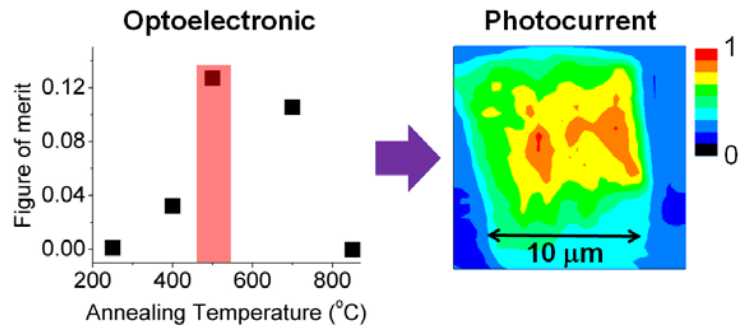

**1. Introduction** Graphene has extraordinary physical properties, making it potentially useful as a material for next generation electronic and optoelectronic devices [1, 2]. It is a semi-metal with extraordinary high charge carrier mobility of 200,000 $cm^2$/Vs, and possesses a high optical transmittance of 98.7 % in the visible range due to the materials atomically thin nature, yet mechanically stable [3-5]. The fundamental bottleneck of its pristine form, however is a lack of intrinsic band gap. Stacking multiple layers of graphene induces a band gap but it is in the range of few tens to a hundred meV [6] thus too small for many of applications. Nanoribbons of graphene with a width range of few tens of nm will also exhibit band gaps (few hundred meV) but their fabrication process generally requires sophisticated instrumentations, and the gap is very sensitive to the width and edge states [7, 8]. Recently, number of reports demonstrated a potential of chemical functionalization as a promising alternative route to gain tunability in graphene electronic structure. More specifically, fluorination and hydrogenation of graphene open a band gap that is as large as one to few eV [9-12]. While these chemical functionalization hold promise toward tunability of graphene electronic structure for practical devices, the functionalization processes generally require vacuum-based system and elevated temperatures (400-600 $^o$C) [9, 10] with an exception of recent efforts using a laser-assisted photochemical method [13, 14]. These process requirements can be disadvantageous for applications such as flexible and organic electronics/optoelectronics, which the low temperature and low cost processing are crucial.

Functionalization using oxygen is an attractive candidate for the purpose. While an atomically controlled functionalization using oxygen gas remains to be challenging [15], inhomogeneous but relatively high oxidization of graphene up to 40 at.% is readily possible *via* simple chemical oxidization of graphite powders at a room temperature (referred as graphene oxide (GO)) [16-18]. In addition, the material properties can be tuned by removing oxygen at nearly room temperature *via* chemical route [16, 17, 19-23]. Further chemical reduction is possible *via* e-beam [24] and laser irradiation [25]. It is well-established from the electrical characterizations that a chemical reduction process can restore semi-metallic property of GO from an insulator [17, 26-29].

In this study, we performed a real-time ultraviolet photoelectron spectroscopy (UPS) of as-synthesized GO upon thermal reduction. A detailed valence band (VB) electronic structure evolution with emphasis on the density of states (DOS) for carbon 2p π bonds and the Fermi level $E_F$ were monitored up to 850 $^o$C in 50 $^o$C steps. As a result, a sharp increase of DOS around the $E_F$ was observed at annealing temperature between 600 and 650 $^o$C indicating a band gap closure [30]. A comparison between the VB electronic structure evolutions and electrical conductivity suggested that annealing temperature of ~500$^o$C would lead to material properties suitable for GO-based optoelectronics. This was demonstrated by a photocurrent map, which an individual GO sheet reduced at ~500$^o$C exhibited a uniform and an order of magnitude enhanced photocurrent over an entire sheet compared to as-synthesized counterpart. Our results pave a pathway for a design of optimized graphene-based optoelectronic devices by providing the key values that are related to electronic structures of GO having different density of chemical functional groups.

**2. Results and Discussion**
**2.1 UPS and electrical conductivity Figure 1**(a) shows a schematic of UPS measurement setup and the procedures used in this study. Uniqueness of our method is a use of pulsed heating sequence to prevent effects of magnetic and electric field on collected photoelectrons. The advantage is that data taken at different temperatures can be compared directly without generally required calibrations. Figure 1(b) shows annealing temperature dependent UPS spectra for range of Fermi level $E_F$ to 5 eV into VB, which provides information on DOS for carbon 2p π bonds (3 eV) and the Fermi level $E_F$ (0 eV) of GO (see Experimental methods for determination of $E_F$). The spectra for annealing temperature of 200 $^o$C and above are shown to exclude the charging effect. After a gradual increase for the π bonds and $E_F$ up to ~600 $^o$C, an unexpected drastic increase was observed between 600 and 650 $^o$C for both cases. In order to gain further insights into the details of the evolution, UPS intensity at 3 eV and 0.3 eV was plotted as a function of annealing temperatures (Figure 1(c)) [30, 31]. 0.3 eV was used instead of 0 eV for $E_F$ merely for its higher intensity with identical evolution behaviors. The results clearly show a drastic increase of DOS at the π bonds and $E_F$ between 600 and 650 $^o$C, which indicates a band gap closure. After the drastic increase between 600 and 650 $^o$C, the π bonds show gradual decrease while the $E_F$ continue to increase at a reduced rate. One possible scenario for the observed trend is a formation of vacancy defects that are reported to occur around the temperature range [32]. Desorption of oxygen groups such as carbonyl and ether in the form of CO or $CO_2$ causes loss of carbon atoms in the basal plane of GO. These losses can lead to a slight decrease of the π bonds above 650 $^o$C. On the other hand, formations of vacancy defects increases the dangling bonds (*i.e.* edge states) in the material system, hence could lead to the observed gradual increase of DOS at $E_F$ [33].

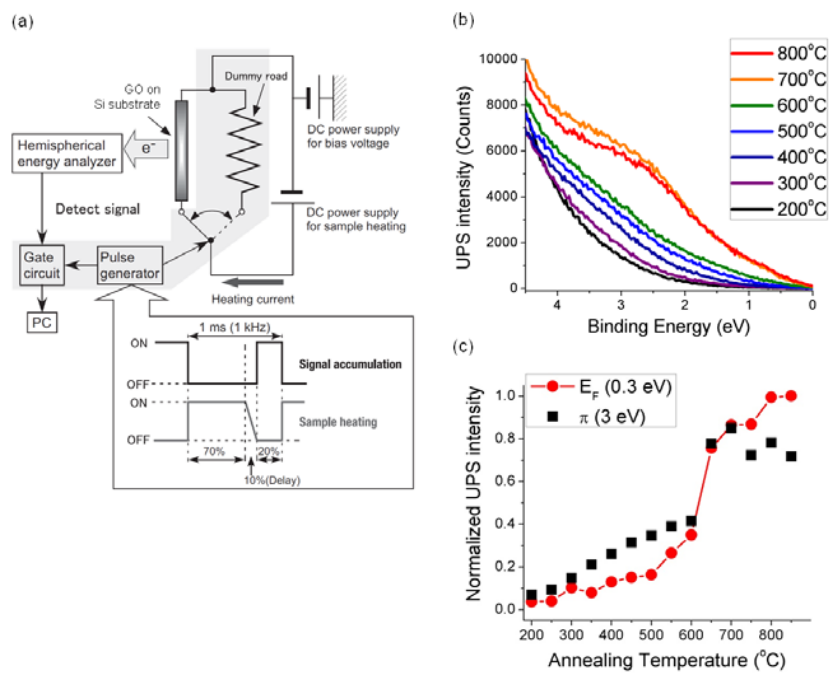

**Figure 1** (a) Schematic of UPS measurement setup and the procedures used in this study. (b) UPS spectra as a function of annealing temperature for energy range between $E_F$ and 5 eV into valence band (VB) of GO. The intensity is in counts and has not been processed. (c) UPS intensity evolution for carbon 2p π bonds and $E_F$. The intensity is normalized for a demonstration purpose to the maximum of each of π bonds and $E_F$.

    The observed drastic increase of DOS for the π bonds and $E_F$ between 600 and 650 °C is unexpected because the temperature range is quite different from what can be expected from well-established knowledge of sharp restoration in electrical conductivity. **Figure 2** (a) shows comparison of evolution between electrical conductivity and UPS intensity as a function of annealing temperature. Difference of annealing temperature which the sharp rise is observed in electrical conductivity and DOS at $E_F$ is clearly demonstrated (purple arrow). One example which the difference would play an important role in is upon a design of GO-based optoelectronics. For an optoelectronic device such as photo-detector, one must consider a balance between electrical conductivity and a possession of band gap. Specifically, a higher electrical conductivity leads to a better device performance but a material should not be metallic as it will lose the photo-response. Therefore, in a material like GO which undergoes semiconductor to semi-metal transitions upon thermal annealing, it is crucial to identify the annealing temperature that leads to material's maximum electrical conductivity but still remain a band gap. Figure 2(b) shows an inverse of normalized UPS intensity at $E_F$ and electrical conductivity as a function of annealing temperature. As UPS intensity at $E_F$ can be used to determine the annealing temperature which a band gap closes (transition to metallic / semi-metal), its inverse could be used as an indication of the materials photo-response. To be more specific, an inverse of UPS intensity drops drastically between 600 and 650 °C for GO, indicating the loss of its optical activity by those annealing temperature range. Based on above mentioned idea, one can consider a figure of merit as shown in Figure 2(c) to obtain quantitative temperature range for the optimal GO-based optoelectronic devices. The figure plots the multiple of two parameters shown in Figure 2(b); electrical conductivity and inverse of UPS intensity at $E_F$. The result suggests that optoelectronic properties can maximize after annealing the GO around ~500 °C.

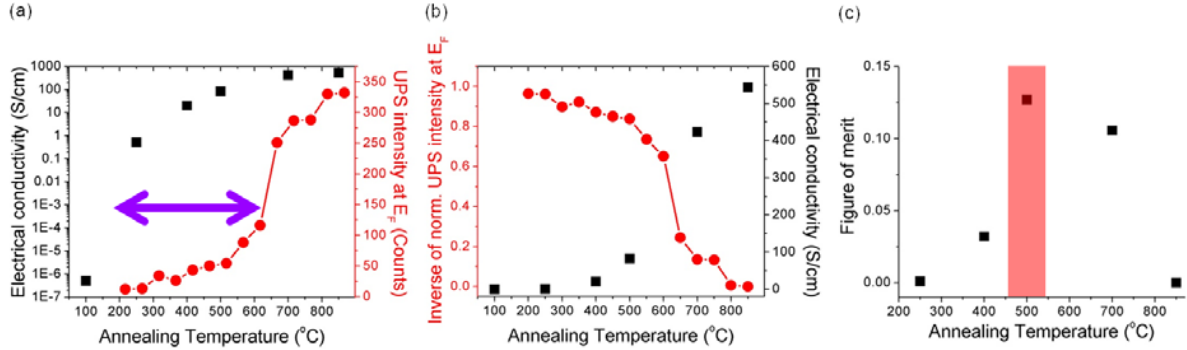

**Figure 2** (a) Comparison between evolution of electrical conductivity (black) and UPS intensity at $E_F$ (red) as a function of annealing temperature. Electrical conductivity is shown in a log scale with S/cm, and the UPS intensity is in a linear scale with counts. (b) Comparison between inverse of UPS intensity at $E_F$ (red) and electrical conductivity (black) as a function of annealing temperature. UPS intensity is normalized to its maximum value, and the conductivity is in S/cm. Both of them are plotted in a linear scale. (c) Multiples of two parameters shown in (b). Pink bar serve as guides for the eyes.

**2.2 XPS and HRTEM** X-ray photoelectron spectroscopy (XPS) and high resolution transmission electron microscopy (HRTEM) were performed to gain insights into a possible origin for the drastic increase of DOS at the π bonds and $E_F$. The simplest scenario for the observed increase is due to the increase of $sp^2$ carbon content around that annealing temperature range. If it was the case, then it could well-explain both increase of DOS at the π bonds and $E_F$. However, based on our XPS analysis as shown in **Figure 3** (a), there is no drastic increase of $sp^2$ carbon at the temperature range between 600 and 650 °C (see Supporting Information for details). Furthermore, the gradual increase of $sp^2$ carbon at the temperature range is consistent with a gradual decrease of O1s. Therefore, XPS results quest for an alternative explanation for the observed DOS increase.

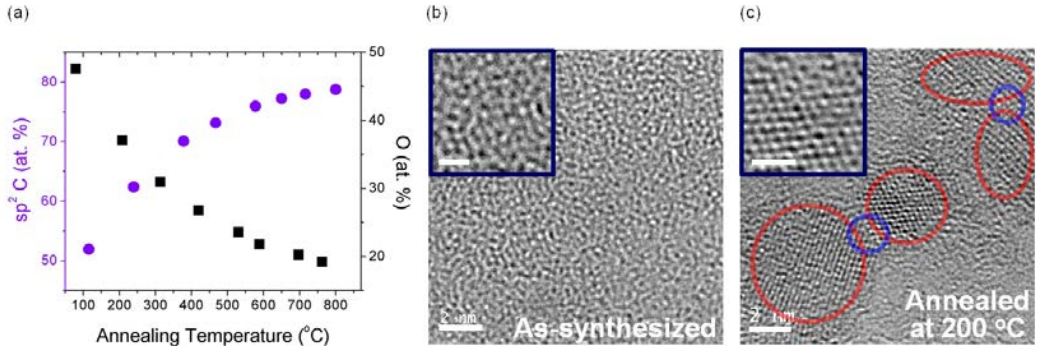

**Figure 3** (a) $sp^2$ carbon (purple) and O1s (black) contents obtained by XPS as a function of annealing temperature. (b) HRTEM of as-synthesized GO. (c) HRTEM image of GO thermally annealed at 200 °C. The red circles indicate regions of $sp^2$ graphene nano-island clusters, and blue circles indicate regions of $sp^2$ graphene percolation chains. Scale bars for (b) and (c) are 2 nm, and 1 nm for insets.

Figure 3 (b) and (c) show the high resolution transmission electron microscopy (HRTEM) image of as-synthesized GO and after annealing at 200 °C, respectively. Due to high density of oxygen functional groups that are inhomogeneously attached to graphene basal plane, no crystalline features are observed for (b). This trend is consistent with literature reporting $sp^2$ carbon clusters / graphene nano-islands in the small domain size of <1 nm [34]. Density variations of graphene nano-islands between the literature and this study are most likely due to minor variations in the synthesis procedures. The observed amorphous nature of as-synthesized GO is the most probable origin of its band gap. Based on the well-studied case of amorphous carbon (a-C), mixture of $sp^2$ and $sp^3$ carbon with $sp^2$ fraction of ~40 at.% (similar to the as-synthesized GO case) is sufficient to explain the gap of few eV [35]. When GO is annealed at 200 °C, graphene nano-islands in the average lateral size of ~2 nm are clearly observed, in contrast to the as-synthesized case (shown in red circles) (Figure 3 (c)). Moreover, a closer look at the images suggests possible percolation starting between graphene nano-islands in the close proximity *via* $sp^2$ carbon links or chains (shown in blue circles). Percolation of graphene nano-islands that is already occurring at 200 °C is expected to proceed further as annealing temperature increases (as some literature has shown up to ~550 °C) [34].

A driving force for the percolation is most likely a combination of increased diffusion rate of oxygen functional groups at elevated temperatures and their energy favorable clustering. As for diffusion rates of oxygen groups, Arrhenius law G=$G_0$ exp(-E/kT), where G is rate of diffusion, $G_0$ is rate constant, and E is migration barrier, indicates that they increase by orders of magnitude at 600 °C compared to the room temperature case. For example, a diffusion rate of hydroxyl group, which is a groups known to remain at 600 °C, increased by as much as 5 orders of magnitude compared to that of room temperature when using the reported diffusion barrier of 0.53 eV [33]. In addition, a tendency of oxygen functional groups in GO to form clusters based on chemical kinetics was recently demonstrated by Kumar *et al.*[36]. This is supported by our HRTEM results on GO, which showed formation of graphene nano-islands (*i.e.* oxygen clusters) after annealing at 200 °C (Figure 3(c)). Although more detailed analysis will be required to conclude on the origin of the drastic increase of DOS at $E_F$ at a temperature range between 600 and 650 °C, our basic materials characterizations that are consistent with literature direct towards enlargement of effective graphene-islands *via* percolation as a possibility.

**2.3 SPCM** In order to demonstrate that optimal annealing conditions for GO optoelectronic devices can be determined based on a combination of DOS at $E_F$ and electrical conductivity, photocurrent of individual GO sheet was obtained. Specifically, two forms of data were obtained using scanned photocurrent microscopy (SPCM) (see Experimental methods for details); a fixed-spot current-time characteristic, and a spatial map [37, 38]. We used annealing temperature of ~500 °C based on figure of merit shown in Figure 2(c). The results indeed proved that high optoelectronic performance can be achieved in devices that were annealed at the temperature range. While as-synthesized GO did not exhibit any photocurrent above the measurement noise level (~1pA), the device annealed at ~500 °C exhibit an order of magnitude enhancement in photocurrent when laser illumination was on the device (9-15 pA, Figure 4 (b)). Moreover, the map showed uniform and enhanced photocurrent over the entire rGO sheet with channel length of as large as > 10 μm (Figure 4(c),(d)).

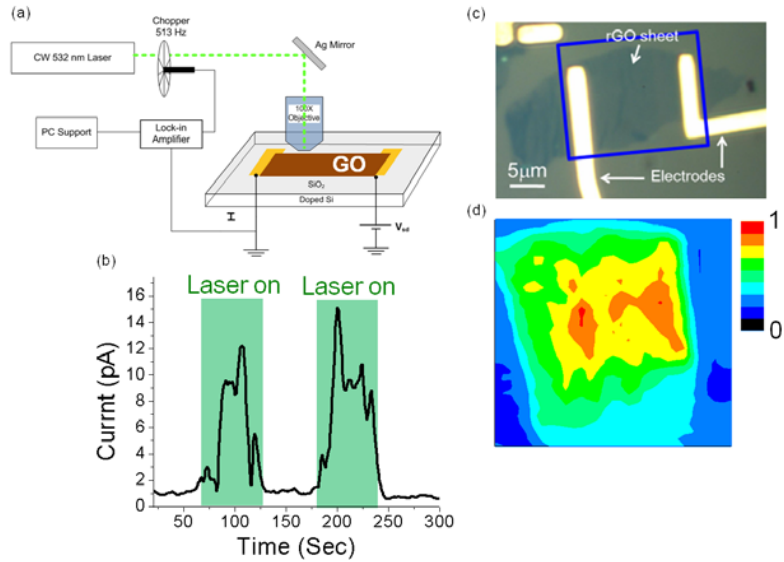

**Figure 4** (a) Schematic of SPCM setup. (b) Photocurrent-time characteristics of GO individual sheet device shown in (c). Fixed-spot laser illumination (λ = 532 nm) was turned on and off periodically as shown with light green bars. (c) Optical microscope image of a device used to obtain (b) and (d). Blue rectangle region indicate the region of (d) SPCM map. Scale bar is 5 μm. (d) SPCM map of individual rGO sheet device annealed at ~500 °C. The size of the map is ~10 μm X 10 μm. The photocurrent is normalized to its maximum value, with red indicating the maximum intensity and black indicating the minimum.

**3. Conclusions** In summary, VB electronic structure evolutions of GO upon thermal reduction is characterized by real-time UPS. DOS around the Fermi level showed a sharp increase at annealing temperature between 600 and 650 °C, indicating a band gap closure. Correlation between evolution of DOS at $E_F$ and chemical / structural properties indicated that a percolation of $sp^2$ carbon as a possible origin. In addition, a combination of UPS and electrical measurements results suggested that annealing GO at a temperature of ~500 °C would lead to the material properties suitable for optoelectronics. This was successfully demonstrated by a uniform and an order of magnitude enhancement in photocurrent map for an individual GO sheet reduced at the same temperature range. Our results provide experimental evidence that demonstrates an importance of studying the electron structures in accordance with other material characterization techniques in achieving optimized device performances, such as for graphene-based optoelectronic devices.

### 4. Experimental Methods

*Synthesis of GO, thin films deposition, and individual sheet device fabrication:* GO was prepared *via* modified Hummer's method [39]. Briefly, graphite powders (Asbury Carbon Inc.) with average flake size of ~500 μm were chemically oxidized for 5 days at room temperature in mixture of sulfuric acid, sodium nitrate, and potassium permanganate at a chemical equilibrium state and exfoliated. After removal of residual metal ions and unexfoliated graphitic particles by continuous washing using deionized (DI) water and repeated centrifugation, respectively, dilute GO suspension in DI water was prepared for thin film deposition. GO films with thickness of ~5 nm were deposited using vacuum filtration on cellulose membrane supports [17], which were then transferred on to $SiO_2$/Si substrates. Cellulose membrane supports were dissolved thoroughly using multiple acetone baths cleaning process, leaving GO films ready for UPS measurements. Deposited GO thin films were also used for XPS, Electrical measurements, and PL.

For the TEM and photocurrent map, synthesized GO aqueous suspension was diluted further and drop-casted onto mesh grids and pre-patterned $SiO_2$/Si substrates, respectively, to obtain isolated individual GO sheets. For photocurrent map, the positions of the monolayer GO individual sheets with respect to the alignment marks were identified under the optical microscope and conventional electron-beam lithography was used to define electrodes on the sheets. The GO sheets were contacted by thermally evaporating Ti/Au (5/30 nm) followed by a lift-off process.

*Real-time UPS measurements:* UPS was performed in an ultrahigh vacuum (UHV) chamber with a base pressure of ~$1 \times 10^{-8}$ Pa. Specimen were gradually heated *in-situ* to ~850 °C in 50 °C increments. At every 50 °C step, temperature was held constant for ~ 700 s while UPS was performed. He-I resonance line (hv=21.22 eV) was used as an excitation source. We used a pulsed heating method to remove artificial effects on the data from magnetic and electric field induced by electric current used for the Joule heating of the specimen (Fig. 1(a)). Mo plate in contact with GO was used as a reference for obtaining Fermi level of the specimen. The specimen was negatively biased (-7 V) to allow efficient collection of secondary electrons near the vacuum level. The energy resolution of the measurement was ~0.10 eV with pass energy of 10 eV. Only the spectra for annealing temperature of 200 °C and above were analyzed in this study due to the possible charging effect for <200 °C.

*ex-situ thermal annealing of GO:* All of the thermal annealing of GO except for UPS was performed *ex-situ*. More specifically, GO films on a substrate was placed in a quartz tube furnace and heated to a targeted temperature in a presence of an inert gas ($N_2$) and kept for 15 minutes.

*Other material characterizations*

*X-ray Photoelectron Spectroscopy (XPS):* XPS measurements were performed with a Thermo Scientific K-Alpha spectrometer. All spectra were taken using Al Ka micro-focused monochromatized source (1486.6 eV) with a resolution of 0.6 eV. The spot size was 400 μm and the operating pressure was $5 \times 10^{-9}$ Pa. The GO thin films were measured on Pt foil and Pt 4f7/2 was taken as reference at 70.98 eV.

*High Resolution Transmission Electron Microscopy (HRTEM):* HRTEM was performed using JOEL JEM-2100F TEM/STEM with double spherical aberration (Cs) correctors (CEOS GmbH, Heidelberg, Germany) to attain high contrast images with a point-to-point resolution of 1.4 Å. The lens aberrations were optimized by evaluating the Zemlin tableau of an amorphous carbon. The residual spherical aberration was almost zero (Cs = -0.8±1.2 μm with 95 % certification). The acceleration voltage was set to 120 kV which is the lowest voltage with effective Cs correctors in the system. The region of interests was focused and then recorded with total exposure of less than 20 s (0.5 s exposure time for the image).

*Electrical Measurements:* Electrical conductivity of GO thin films was obtained from sheet resistance measured using standard Van der Pauw method. Keithley 2400 sourcemeter was connected to four indium contacts soldered on each corner of the GO thin films. The film size used for the measurements was typically 15 mm x 15 mm and area of indium contacts was kept less than 1 mm x 1 mm. We have confirmed that the sheet resistance measured by Van der Pauw method and conventional two probe method with Au contacts (Length: 20 μm, Width: 400 μm) results in similar values.

*Scanning Photocurrent Microscopy (SPCM):* Green diode pumped solid state laser (λ = 532 nm) was coupled through a confocal microscope *via* 100x Olympus objective with 0.9 numerical aperture to scan over the specimen to form a photocurrent map [37]. The laser spot size was ~400 nm. The laser power was kept less than 1 mW to prevent possible thermal reduction as well as damage to the GO / rGO devices. Synchronous detection was achieved *via* SR-830 lock-in amplifier (Stanford Research Systems) using the reference frequency input from the chopper (~370 Hz). The photocurrent signal was first amplified by a SR-570 current amplifier (Stanford Research Systems) and the AC component of the output voltage signal was used as direct input to the lock-in amplifier. Both the amplitude *R* and the phase φ of the photocurrent were monitored. Correlated PL obtained for the same position of SPCM map was detected using an avalanche photodiode, allowing determination of the exact location of individual monolayer GO sheet device and the contact electrodes. 1 V was applied between the electrodes for the measurements performed in this study. A center region of the device was used for fixed-spot measurements.

**5. Supporting Information** Supporting Information on Atomic force microscopy (AFM) of GO sheets and XPS performed in this study are available online is available from the Wiley Online Library or from the author.

**Acknowledgements** The authors acknowledge T.Kaga and S.Takabayashi of Tohoku University, Japan, and E.Cheng and D.Voiry of Rutgers University for the experimental support. The authors also acknowledge Asbury Carbon, NJ for generously supplying the starting graphite powders as a part of their U.S. national laboratory supporting program. H.Y. and M.C. acknowledge Donald H. Jacobs' Chair funding from Rutgers University. H.Y. acknowledges the Laboratory Directed Research and Development (LDRD) Director's Postdoctoral Fellowship of LANL, and the Japanese Society for the Promotion of Science (JSPS) Postdoctoral Fellowship for Research Abroad for financial support. This work was performed under the Cooperative Research Program of the "Network Joint Research Center for Materials and Devices" by the Ministry of Education, Culture, Sports, Science and Technology (MEXT), Japan. The XPS measurements using synchrotron radiation were performed at BL23SU in SPring-8 under "Nano-net Project" of the Japan Synchrotron Research Institute (JASRI) and Japan Atomic Energy Agency (JAEA) (proposal Nos. 2010A3874, 2010B3879, and 2014B3874). The research was also supported by the LDRD Program and performed, in part, at the Center for Integrated Nanotechnologies, an Office of Science User Facility operated for the U.S. Department of Energy (DOE) Office of Science. Los Alamos National Laboratory, an affirmative action equal opportunity employer, is operated by Los Alamos National Security, LLC, for the National Nuclear Security Administration of the U.S. Department of Energy under contract DE-AC52-06NA25396.

Supporting Information

# Valence band electronic structure evolution of graphene oxide upon thermal annealing for optoelectronics

*Hisato Yamaguchi\*, Shuichi Ogawa, Daiki Watanabe, Hideaki Hozumi, Yongqian Gao, Goki Eda, Cecilia Mattevi, Takeshi Fujita, Akitaka Yoshigoe, Shinji Ishizuka, Lyudmyla Adamska, Takatoshi Yamada, Andrew M. Dattelbaum, Gautam Gupta, Stephen K. Doorn, Kirill A. Velizhanin, Yuden Teraoka, Mingwei Chen, Han Htoon, Manish Chhowalla, Aditya D. Mohite\* and Yuji Takakuwa\**

**Contents**
- S1   Atomic force microscopy (AFM)
- S2   X-ray photoelectron spectroscopy (XPS)

*S1 Atomic force microscopy (AFM)*

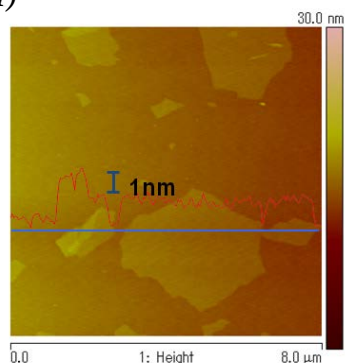

Figure S1. AFM image of typical graphene oxide (GO) sheets.

*S2 X-ray photoelectron spectroscopy (XPS)*

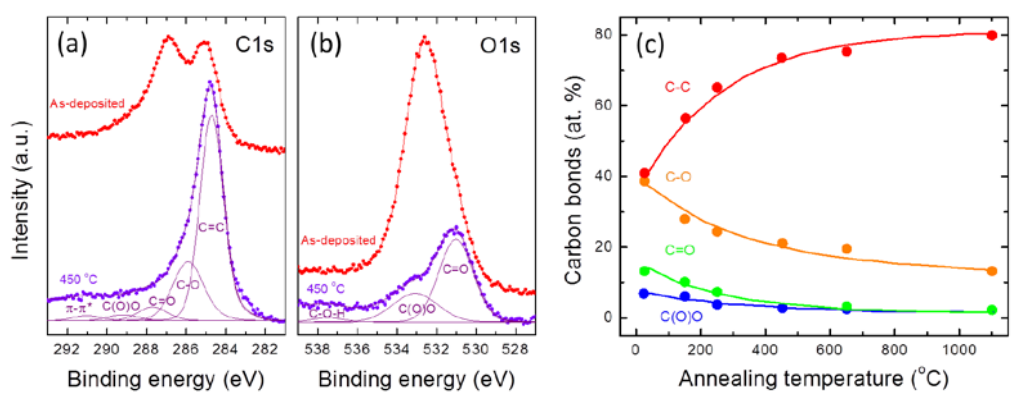

Figure S2. XPS spectra of graphene oxide (GO) films corresponding to **(a)** C1s and **(b)** O1s peaks for as-deposited and annealed at 450 °C. Deconvoluted peaks are shown for 450 °C annealed cases. **(c)** Evolution of C-C (sp$^2$) bonds in respect to oxygen containing bonds (C-O, C=O, and C(O)O)) in at. % as a function of annealing temperature (re-plotted using the data from our previous work Ref.[S1]). Refer to our previous study for details on assignments of deconvoluted peaks [S1].

**Reference**
[S1] C. Mattevi, G. Eda, S. Agnoli, S. Miller, K. A. Mkhoyan, O. Celik, D. Mastrogiovanni, G. Granozzi, E. Garfunkel, M. Chhowalla, Advanced Functional Materials **19**, 2577 (2009).